\documentclass[aps,floatfix,superscriptaddress]{revtex4}
\usepackage{amsmath,amssymb,amsfonts,graphics,graphicx,dcolumn,bm,enumerate}
\usepackage{comment,natbib}
\usepackage[ddmmyyyy,hhmmss]{datetime}
\usepackage{multirow,color}

\newcommand{\cmp}
{\affiliation{Condensed Matter Physics Division, 
Saha Institute of Nuclear Physics, 1/AF Bidhannagar, Kolkata 700064, India.}}
\newcommand{\isi}
{\affiliation{Economic Research Unit, Indian Statistical Institute, 203 B. T. 
Road, Kolkata 700108, India.}}
\newcommand{\aalto}
{\affiliation{Department of Computer Science, TUAS Building, Otaniementie 17, 
02150
Espoo,
Finland.}}
\newcommand{\hok}
{\affiliation{Graduate School of Information Science \& Technology,  Hokkaido 
University, N14-W9, Kita-ku, Sapporo 060-0814, Japan.}}

\begin{document}
% \title{The k-index for the exponential, power-law, 
% uniform, log-normal distributions,  
% and those mixture with a crossover}
\title{Social inequality: from data to statistical physics modeling}

\author{Arnab Chatterjee}%
\email[Email: ]{arnabchat@gmail.com}
\cmp
\author{Asim Ghosh}
%\email[Email: ]{asim.ghosh@saha.ac.in}
\cmp \aalto
\author{Jun-ichi Inoue\footnote{Dedicated to the memory of Prof. J.-I. 
Inoue}}
%\email[Email: ]{jinoue@cb4.so-net.ne.jp, j_inoue@complex.ist.hokudai.ac.jp}
\hok 
\author{Bikas K. Chakrabarti}%
%\email[Email: ]{bikask.chakrabarti@saha.ac.in}
\cmp \isi

% \date{\today}

%%%%%%%%%%%%%%%%%%%%%%%%%%%%%%%%%%%%%%%%%%%%%%%%%%%%%%%
%%                                                                               Abstract                                                                                             %%
%%%%%%%%%%%%%%%%%%%%%%%%%%%%%%%%%%%%%%%%%%%%%%%%%%%%%%%
\begin{abstract}
Social inequality is a topic of interest since ages,
and has attracted researchers across disciplines to ponder over it origin, 
manifestation, characteristics, consequences, and finally, the question of how 
to cope with it. It is manifested across different strata of human existence, 
and is quantified in several ways.
In this review we discuss the origins of social inequality, the historical and 
commonly used non-entropic measures such as Lorenz curve, Gini index and the 
recently introduced $k$ index. We also discuss some analytical tools that aid in 
understanding and characterizing them.
Finally, we argue how statistical physics modeling helps in reproducing
the results and interpreting them.
\end{abstract}

\maketitle 

%%%%%%%%%%%%%%%%%%%%%%%%%%%%%%%%%%%%%%%%%%%%%%%%%%%%%%%%%
\section{Introduction}
% In humans,  social interactions are mostly complex and nonlinear.
Repeated social interactions produce spontaneous variations 
manifested as inequalities at various levels. 
With the availability of huge amount of empirical data for a plethora of 
measures of human social interactions makes it possible to uncover the 
patterns and look for the reasons behind socio-economic inequalities.
Using tools of statistical physics, researchers are bringing in 
knowledge and techniques from various other disciplines~\cite{lazer09}, e.g., 
statistics, applied mathematics, information theory and computer science 
to have a better
understanding of the precise nature (both spatial and temporal) and the origin 
of socio-economic inequalities that is prevalent in our society.

%%%%%%%%%%%%%%%%%
Socio-economic 
inequality~\cite{arrow2000meritocracy,stiglitz2012price,neckerman2004social,
goldthorpe2010analysing,chatterjee2015sociorev} 
is concerned with the existence of unequal opportunities and rewards for 
various social positions or statuses in a society.
Structured and recurrent patterns of unequal distributions of goods, wealth, 
opportunities, and even rewards and punishments are the key features, and
measured as \textit{inequality of conditions},
and \textit{inequality of opportunities}.
The first one refers to the unequal distribution of income,
wealth and material goods, while the latter refers to the unequal distribution 
of `life chances' across individuals, as is reflected in 
education, health status, treatment by the  criminal justice system, etc.
Socio-economic inequality is held responsible for conflict, war, crisis, 
oppression, criminal activity, political unrest and instability, and
affects economic growth  indirectly~\cite{hurst1995social}.
Usually, economic inequalities have been studied in the context of income 
and wealth~\cite{yakovenko2009colloquium,chakrabarti2013econophysics,
aoyama2010econophysics}. However, it is also measured for a plethora of
quantities, including energy consumption~\cite{lawrence2013global}.
%%%%%%%%%%%%
The inequality in 
society~\cite{piketty2014inequality,Cho23052014,Chin23052014,Xie23052014}
is an issue of current focus and immediate global interest, bringing together
researchers across several disciplines -- economics and finance, sociology, 
demography,  statistics along with theoretical physics (See e.g., 
Ref.~\cite{chakrabarti2013econophysics,Castellano:2009,Sen:2013}).

Socio-economic inequalities are quantified in numerous ways. 
The most popular measures are absolute, as quited with a single number, 
in terms of indices, e.g., Gini~\cite{gini1921measurement}, 
Theil~\cite{theil1967economics}, Pietra~\cite{eliazar2010measuring} indices.
The alternative measure approach is relative,
using probability distributions of various quantities, but the most of the 
previously mentioned indices can be computed once one has the knowledge of the 
distributions. 
What is usually observed is that most quantities 
display broad distributions, usually lognormals, power-laws or their 
combinations.
For example, the distribution of income is usually an exponential followed by a 
power law~\cite{chakrabarti2013econophysics,druagulescu2001exponential}.

\begin{figure}[t]
\begin{center}
\includegraphics[width=10.0cm]{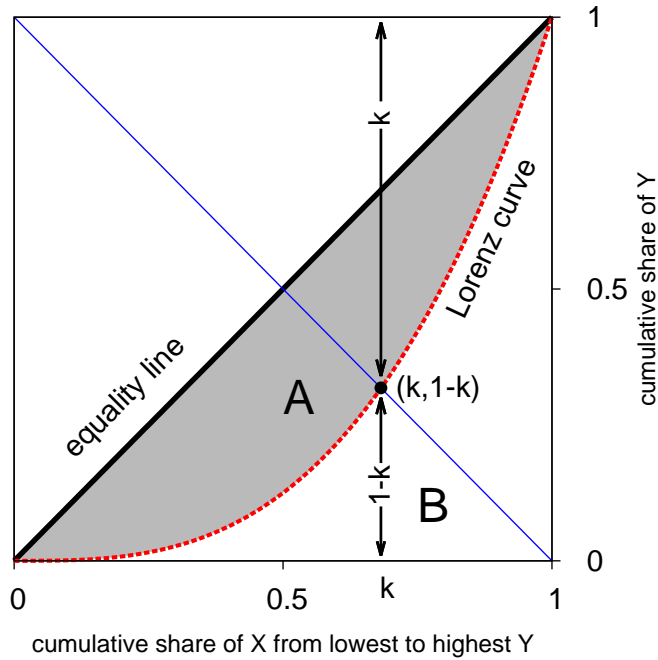}
\end{center}
\caption{
Schematic representations for  
Lorenz curve, Gini index $g$ and $k$ index.
The dashed line depicts the Lorenz curve and the solid line represents 
perfect equality.
The area enclosed between the equality line and the Lorenz curve is $A$ and 
that below the Lorenz curve is $B$.
The Gini index is given by $g=A/(A+B)$.
The $k$ index is given by the abscissa of the intersection point of the Lorenz 
curve and $Y=1-X$ (adapted from Ref.~\cite{inoue2015measuring}).
}
\label{fig:fg_pic_Lorenz}
\end{figure}

In one of the popular methods of measuring inequality, one has to 
consider the Lorenz curve~\cite{Lorenz}, which is a function that represents 
the cumulative proportion $X$ of ordered (from lowest to highest) individuals  
in terms of the cumulative proportion of their sizes $Y$.
$X$ can represent income or wealth of individuals.
But it can as well be citation, votes, city population etc. of articles, 
candidates, cities etc. respectively.
The Gini index ($g$) is defined as the ratio between the area 
enclosed between the Lorenz curve and the equality line, to that below the 
equality line.
If the area between 
(i) the Lorenz curve and the equality line is represented as $A$, and 
(ii) that below the Lorenz curve as $B$  (See Fig.~\ref{fig:fg_pic_Lorenz}),
the Gini index is $g=A/(A+B)$.
It is an useful measure to quantify socio-economic inequality.
Ghosh et al.~\cite{ghosh2014inequality}
recently introduced the `$k$ index' (where `$k$' symbolizes for the extreme 
nature of social inequalities in Kolkata), which is 
defined as the fraction $k$ such that  $(1-k)$ fraction 
of people or papers possess $k$ fraction of income or citations 
respectively~\cite{inoue2015measuring}.

%%%%%%%%%%%%%%%%%%%%%%%%%%%%%%%%%%%%%%%%%%%%%%%%%%%%%%%%
%%%%%%%%%%%%%%%%%%%%%%%%%%%%%%%%%%%%%

When the probability distribution is described using an appropriate parametric 
function, 
one can derive these inequality measures as a function of those parameters 
analytically. 
In fact, several empirical evidence have been reported to show 
that the distributions can be put into a finite number of types. 
Most of them turn out to be a of mixture of two distinct parametric 
distributions with a single crossover point~\cite{inoue2015measuring}.  

This review is organized as follows:
Sec.~\ref{sec:evolution} discusses the evolutionary view of socio-economic
inequalities and Sec.~\ref{sec:Generic} discusses the basic and most popular 
quantities which are used for measuring inequalities.
Next, in Sec.~\ref{sec:results} we discuss the most realistic scenario in which 
probability distributions have to fitted to more than a single function, and 
how to measure inequalities from them. Further, we discuss if inequalities are
natural, in context of the recent works of Piketty in Sec.~\ref{sec:natural},
followed by a section on statistical physics modeling in Sec.~\ref{sec:spmodel}.
Finally, we summarize in Sec.~\ref{sec:summary}.

%%%%%%%%%%%%%%%%%%%%%%%%%%%%%%%%%%%%%%%%%%%%%%%%%%%%%%%%%%%%%%%%%%%%%%%%%

\section{Evolutionary view of socio-economic inequality}
\label{sec:evolution}
The human species is known to have lived their life as hunter gathers for more 
than 90\% of its existence. 
It is widely believed that those early societies were egalitarian, 
still seen in the lifestyle of various tribes like the !Kung people of the 
Kalahari~\cite{pennisi2014our}. 
Early hunter-gatherer societies are even believed to have 
championed sex equality~\cite{dyble2015sex}.
Traditionally having a few possessions, they were semi nomadic in the sense 
they were moving periodically.
Having hardly mastered farming and lived as small groups, the mere survival 
instinct was driving them to overlook individual interests.
They were sharing what they had, so that all of their group members were 
healthy and strong, be it food, weapons, property, or 
territory~\cite{buchanan2014more}.
With the advent of agricultural societies, elaborate hierarchies were created, 
with less stable leadership in course of time. These evolved into clans or 
groups led mostly by family lines, which eventually developed as kingdoms.
In these complex scenarios, new strategies for hoarding surplus
produce of agriculture or goods were adapted by the chiefs or kings, 
predominantly for survival in times of need, and this lead to the concentration 
of wealth and power (see Ref.~\cite{Chatterjee-2007}  for models with savings). 
Along with the advancement of technologies, intermediate mechanisms helped in 
wealth multiplication.
This completed the transition from egalitarianism to societies having 
competition and the inequality
paved way for the growth of chiefdoms, states and industrial 
empires~\cite{pringle2014ancient,chatterjee2015sociorev}.

%%%%%%%%%%%%%%%%%%%%%%%%%%%%%
%%%%%%%%%%%%%%%%%%%%%%%%%%%%%%%%%%%%%%%%%%%%
\section{Basics and generic properties of inequality measures}
\label{sec:Generic}
%%%%%%%%%%%%%%%%%%%%%%%%%%%%%%%%%%%%%%%%%%%%
In this section, we formally introduce the measures to quantify 
the degree of social inequality, namely, 
Lorenz curve, Gini index and $k$ index.

The Lorenz curve shows the 
relationship between 
the cumulative distribution and 
the cumulative first moment of $P(m)$:
%%%
\begin{equation}
 X (r ) =  \int_{m_{0}}^{r}P(m)dm,\,\,\,
 Y (r ) =  
 \frac{\int_{m_{0}}^{r}mP(m)dm}
 {\int_{m_{0}}^{\infty} mP(m)dm}.
 \label{eq:def_Lorenz}
 \end{equation}
 %%%
 The set $\left(X(r ),Y(r )\right)$ defines the Lorenz curve, 
 assuming $P(m)$ to be defined in $[m_{0},\infty)$. 
 Fig.~\ref{fig:fg_pic_Lorenz} shows the typical 
behavior of Lorenz curve.
The Lorenz curve gives  
the cumulative proportion $X$ of ordered individuals  (from lowest to highest) 
holding the cumulative proportion  $Y$ of wealth.
Lorenz curve, Gini index etc. were historically introduced in the context of 
income/wealth. So, let us call  $X$ as individuals and  $Y$ as `wealth',
but in principle the attributes $X$ and $Y$ can be any of the combinations 
like article/citations, candidate/vote, city/population, student/marks,
company/ employee etc.
Hence,  when all individuals take the same amount 
of wealth, say $m^{\prime}$, we have
$P(m)=\delta (m-m^{\prime})$, with $ m_{0} <m^{\prime} <\infty$,
and one obtains 
%%%%
\begin{equation}
X (r ) = \int_{m_{0}}^{r} \delta (m-m^{\prime})dm =\Theta (r-m^{\prime}), \;\;
%%%
Y(r ) = \frac{\int_{m_{0}}^{r} m\delta (m-m^{\prime})dm}
{\int_{m_{0}}^{\infty}m\delta(r-m^{\prime})dm}
=\frac{m^{\prime}\Theta (r-m^{\prime})}{m^{\prime}}=X (r ). 
\end{equation}
%%%%%%%%%%%%%
where 
$\Theta (x)$ is a step function defined by 
%%%
\begin{equation}
\Theta (x) = 
\left\{
\begin{array}{cc}
1, & \quad \quad x \geq 1, \\
0, & \quad \quad x < 1.
\end{array}
\right.
\end{equation}
Thus $ Y=X$ 
 is the `perfect equality line' (see Fig.~\ref{fig:fg_pic_Lorenz}),
 where $X$ fraction of people takes $X$ fraction of wealth in the society.
On the other extreme, 
if the total wealth in the society of $N$ persons is concentrated to 
a few persons,  
%%%%%%
$P(m)=(1-\varepsilon) \delta_{m,0}+\varepsilon \delta_{m,1}$,
%%%%%%%%%%% 
with $\varepsilon \sim \mathcal{O}(1/N)$ and 
with the total amount of wealth is normalized to $1$, 
we get $X(r )=1-\varepsilon +\varepsilon \delta_{r,1}$ and 
$Y(r )=\delta_{r,1}$. 
Hence, $Y=1$ iff  $X=r=1$ and $Y=0$ otherwise, 
and 
the Lorenz curve is given as 
`perfect inequality line' $Y=\delta_{X,1}$
where $\delta_{x,y}$ is a Kronecker's delta (see Fig.~\ref{fig:fg_pic_Lorenz}). 

For a given Lorenz curve, the Gini index is defined by
twice of area between the curve $(X(r ),Y(r ))$ and perfect equality line 
$Y=X$. 
(shaded part  `{\sffamily A}' in Fig.~\ref{fig:fg_pic_Lorenz}). 
It reads 
%%%
\begin{equation}
g= 2 \int_{0}^{1}(X-Y)dX = 
2 \int_{r_{0}}^{\infty}(X(r )-Y(r ))\frac{dX}{dr}dr,
\label{eq:def_G}
\end{equation}
%%%
where  $X^{-1}(0)=r_{0}, X^{-1}(1)=\infty$ should hold. 
Graphically  (see Fig.~\ref{fig:fg_pic_Lorenz}), the Gini index is the ratio of 
the two areas 
(`{\sffamily A}' and `{\sffamily B}'), $g={\rm A}/({\rm A}+{\rm B})$. 
Thus, the Gini index $g$ is zero for perfect equality and unity for 
perfect inequality. 
The Gini index may be evaluated analytically when 
the distribution of population is obtained in a parametric way.

The recently introduced $k$ index is
the value of $X$-axis for 
the intersection between 
the Lorenz curve and a straight line $Y=1-X$. 
For the solution of equation
$X(r )+Y(r )=1$,
%%%
say $r_{*}=Z^{-1}(1), Z(r )\equiv X(r ) + Y(r )$, 
the $k$ index is given by 
%%%
$k =X(r_{*})$.
%%%%%
Thus, the value of $k$ index indicates that
$k$ fraction of people shares totally $(1-k)$ fraction of the wealth. 
Hence, the $k$ index equals $1/2$ for perfectly equal society, 
and $1$ for perfectly unequal society.
This is obviously easier to estimate by eyes 
in comparison with the Gini index (shaded area {\sffamily A} in 
Fig.~\ref{fig:fg_pic_Lorenz}). 

Apart from these indices,
Pietra's $p$ index \cite{eliazar2010measuring} and $m$ or median index 
\cite{eliazar2014socialinequality} has been used as inequality measures, and 
can be derived from the Lorenz curve. 
The $p$ index is defined
by the maximal vertical distance between the Lorenz curve and the line of 
perfect equality $Y=X$,
while the  $m$ index is given by $2m-1$ 
for the solution of $Y(m)=1/2$.

%%%%%%%%%%%%%%%%%%%%%%%%%%%%%%%%%%%%%%%%%%%%%%%%%%%%%
\section{Results for mixture of distributions}
\label{sec:results}
%%%%%%%%%%%%%%%%%%%%%%%%%%%%%%%%%%
It is rather easy to perform analytic calculations for the $g$ and $k$ indices  
when the distribution 
of population are described by parametric distributions such as 
a uniform, power law and lognormal distributions. 
It is quite common to find that the 
probability distributions of quantities like wealth, income, votes, citations 
etc. fit to more than one theoretical function depending on the range.
Formally, 
$P(m) = F_{1}(m)\theta(m,m_{\times})+
F_{2}(m)\Theta (m-m_{\times})$, with 
$\theta (m,m_{\times}) \equiv \Theta (m)-\Theta (m-m_{\times})$, where  
$m_{\times}$ is the crossover point.
The functions $F_1(m)$ and $F_2(m)$ are suitably normalized
and computed for their continuity at $m_{\times}$. 
In such cases, one can also develop a 
framework~\cite{inoue2015measuring}, using which 
it is reasonably straightforward to calculate Lorenz curve, Gini and $k$-index.

%%%%%%%%%%%%%%%%%%%%%%%%%%%%%%%%
\subsection{Empirical data}
\label{subsec:empirical}
\begin{figure}[t]
 \includegraphics[width=17.0cm]{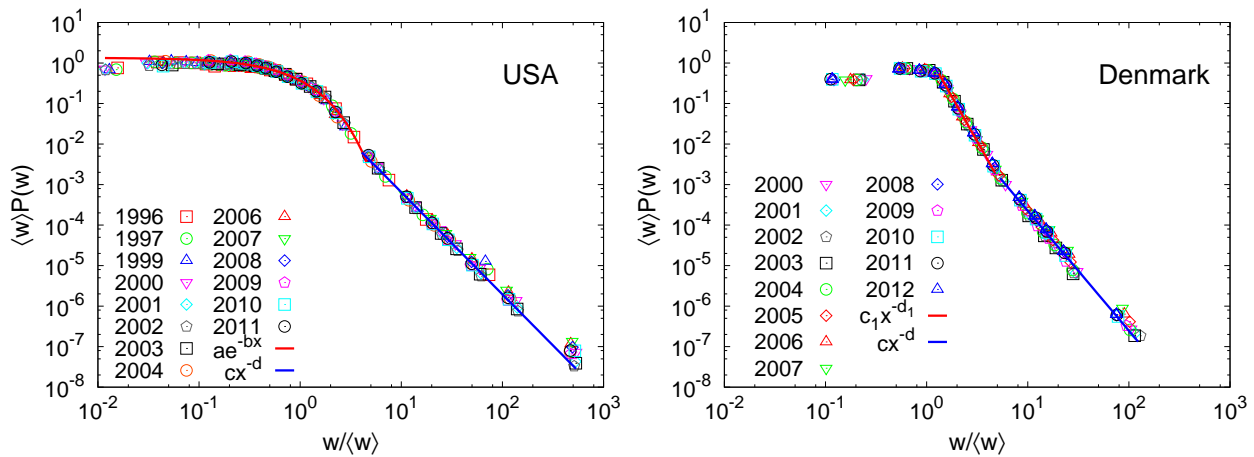}
%%%%%%%%%%%%%%%%%%%%%%%%
\caption{Left: Income distribution $p(w)$ for USA~\cite{IRS} 
for a few years, the data is rescaled by the average income $\langle w \rangle$.
The lower income range fits to an exponential form $a\exp(-bw)$ with $b=1.3$ 
while the high income range fits to a power law decay $c w^{-d}$ with exponent 
$d=2.52$.
Right: Same for Denmark~\cite{Denmarkstat}. The middle and upper ranges fit to 
different power laws $cw^{-d}$ with $d=2.96$ and $c_1 w^{-d_1}$ with $d_1=4.41$.
Taken from Ref.~\cite{inoue2015measuring}.
}
\label{fig:income}
\end{figure}

%%%%%%%%%%%%%%%%
The most well studied data in the context of socio-economic inequality is that 
of income.
Incomes are re-calculated from the income tax data reported in the Internal 
Revenue Service (IRS)~\cite{IRS} of USA for 1996-2011. This data is used to 
compute the probability distribution $P(w)$ of income $w$ for each year.
Similar data from Denmark were used from the years 2000-2012~\cite{Denmarkstat}.
The $g$ index is found to be around $0.54-0.60$ for USA and $0.34-0.38$ for 
Denmark, while the $k$-index is around $0.69-0.71$ for USA and $0.65-0.69$ for 
Denmark~\cite{inoue2015measuring}.
Fig.~\ref{fig:income} shows both these data sets.
%%%%%%%%%%%%%%%%%%%%%%%%%%%%%
%%%%%%%%%%%

The data related to several  socio-economic measures were analyzed. 
The (i) voting data of proportional elections from a few 
countries~\cite{chatterjee2013universality})., 
(ii) citations of different science journals and institutions, 
collected from ISI Web of 
Science~\cite{ISI},
and (iii) population data for city sizes for Brasil~\cite{Brasil}, 
municipalities of Spain~\cite{Spain} and Japan~\cite{Japan}
were analyzed.
The data showed broad distributions of the above quantities, which
well-fitted to (see Ref.~\cite{inoue2015measuring} for details)
(a) a single lognormal, 
(b) a single lognormal with a power law tail, 
(c) uniform with a power law tail, 
(d) uniform with a lognormal tail, 
(e) a mixture of power laws, 
(f) a single power law with a lognormal tail. 

The inequality indices computed using a combination of fitting functions 
were compared with that computed directly from the empirical data.

%%%%%%%%%%%%%%%%%%%%%%%%%%%%%%%%%%%%%%%%%%%
\section{Is wealth and income inequality natural?}
\label{sec:natural}
One is often left wondering whether inequalities in wealth and income are 
natural. It has been shown using models~\cite{chatterjee2007economic} and their 
dynamics that certain minimal dynamics over a completely random exchange 
process and subsequent entropy maximization produces broad distributions.
Piketty~\cite{piketty2014capital} argued recently that inequality in wealth 
distribution is quite natural.
He pointed out that before the great wars of the early 20th century, the 
strong skewness of wealth distribution was prevailing as a result of a certain 
`natural' mechanisms. 
The Great Depression that followed after the two great World Wars have helped 
in dispersion of wealth. This in turn brought the prevailing extreme inequality 
under check and subsequently gave rise to a sizable middle class.
After analyzing very accurate data, he concluded that the world is currently 
`recovering' back to this `natural state', which is happening due to capital 
ownership driven growth of finance~\cite{piketty2014inequality}, which has been 
dominant over a labor economy, and this is simply the result of the type of 
institution and policies that are adopted by the society.
His work raises quite fundamental issues concerning bot only economic 
theory but also the future of capitalism. It points out the large increases in 
the wealth/output ratio. According to standard economic theory, such increases  
are attributed to the decrease in the return to capital and an increase in 
the wages. However, the return to capital has not diminished, while the wages 
have. He has also prescribed the following: higher capital-gains and inheritance 
taxes, higher spending towards access to education, tight enforcement of 
anti-trust laws, corporate-governance reforms that restrict pay for the 
executives, and finally, the financial regulations which have been an 
instrument for banks to exploit the society. It is anticipated that all of 
these might be able to help reduce inequality and increase equality of 
opportunity. There is further speculation that this might be able to 
restore the shared and quick economic growth that characterized the 
middle-class societies in the mid-twentieth century.

%%%%%%%%%%%%%%%%%%%%%%%%%%%%%%%%%%
\section{Statistical physics modeling}
\label{sec:spmodel}
%%%%%%%%%%%%%%%%%%%%%%%%%%%%%%%%%
One of the most efficient ways to model evolution of systems to broad 
distributions showing strong inequality, is by using the toolbox of
statistical physics. Microscopic and macroscopic modeling helps
in imitating real socio-economic systems.

There is a whole body of empirical evidence supporting the fact that a number 
of social phenomena are characterized by emergent behavior out of the 
interactions of many individual social components. Recently, the growing 
community of researchers have analyzed large-scale social dynamics to 
uncover certain `universal patterns'. There has also been an attempt to propose 
simple microscopic models to describe them, in the same spirit as the 
minimalistic models commonly used in statistical 
physics~\cite{Castellano:2009,Sen:2013}.

\subsection{Income \& wealth distribution}
In case of wealth distributions, the popular models are chemical kinetics 
motivated Lotka-Volterra models~\cite{Levy:1997,Solomon:2002,Richmond:2001},
polymer physics inspired models~\cite{Bouchaud:2000} and most importantly, 
models inspired by kinetic theory of 
gases~\cite{Dragulescu:2000,Chakraborti:2000,Chatterjee:2004,Chatterjee-2007} 
(see Ref.~\cite{chakrabarti2013econophysics} for details).
The two-class structure~\cite{yakovenko2009colloquium} of the income 
distribution (exponential dominated low income and power law tail in the high 
incomes) is well understood to be a result of very different dynamics of 
the two classes.
The bulk is described by a process which is more of a random 
kinetic exchange~\cite{Chakraborti:2000,Dragulescu:2000}, producing a 
distribution dominated by an exponential functional form.
The dynamics is very simple, as described in the kinetic theory of 
gases~\cite{saha1958treatise}.
The minimal modifications that one can introduce are additive or 
multiplicative terms.

Processes creating inequality involving uniform retention 
rates~\cite{angle2006inequality} or equivalently, 
savings~\cite{Chakraborti:2000}  produce Gamma-like distributions.
These models are defined as a microcanonical ensemble, with fixed number of 
agents and wealth. Here, the wealth exchanging \textit{agent} retains a certain 
fraction (termed as `saving propensity') of what they had before each trading 
process and randomly exchanges the rest of the wealth. When agents are 
assigned with the same value of the `saving propensity' (as in 
Ref.~\cite{Chakraborti:2000}), it could not produce a broad distribution of 
wealth. What is important to note here is that the richest follow a different 
dynamic from the poor and thus heterogeneity in the saving behavior plays a 
crucial role. So, to obtain the power law distribution of wealth for the 
richest, one needs to simply consider that each agent is different in terms of 
how much fraction of wealth they will save in each 
trading~\cite{Chatterjee:2004},
which is a very natural ingredient to assume, because it is quite likely that 
agents in a trading market think very differently from one another.
In fact, with this small modification, one can explain the entire range of the 
wealth distribution~\cite{Chatterjee-2007}.
These models, moreover, can show interesting characteristics if the exchange 
processes and flows are made asymmetric, e.g., put on directed 
networks~\cite{chatterjee2009kinetic}.
A plethora of variants of these models, results and analyses find 
possible applications in a variety of trading 
processes~\cite{chakrabarti2013econophysics}.

\subsection{Cities \& firms}
City~\cite{Zipf:1949} and firm sizes~\cite{axtell2001zipf} consistently 
exhibit broad distributions with power law tails for the largest sizes,
commonly known to be Zipf's law.

Gabaix~\cite{gabaix1999zipf1} showed that if cities grow randomly at the same 
expected growth rate and the same variance (Gibrat's law~\cite{Gibrat:1931}), 
the limiting distribution converges to the Zipf's law.
He proposed that growth `shocks' are random and they impact utilities in both 
positive and negative way.
A similar approach resulted in diffusion and multiplicative 
processes~\cite{zanette1997role}. Shocks were also used to immitate sudden 
migration~\cite{marsili1998interacting}.
Simple economics arguments demonstrated that expected urban growth rates were 
identical irrespective of city sizes and variations were random normal 
deviates, resulting Zipf law with exponent unity.

\subsection{Consensus}
Consensus in social systems is an interesting topic, due to its dynamics.
The dynamics of agreement and disagreement in a `society' 
is complex, and statistical physicists working on opinion dynamics have been 
brave enough to model opinion states in a population and their dynamics 
that determine the transitions within such states.
A huge body of old and recent literature~\cite{Sen:2013,Castellano:2009}
discusses models that explain various social phenomena and the observed
inequality in such instances of consensus formation.

\subsection{Bibliometrics}
The increasing amount of data produced from bibliometric tools have led to a 
better understanding of how researchers and their publications `interact' with 
one another in a `social system' consisting of articles and researchers.
The patterns of citation distribution and growth are now well studied,
and some of the most successful models have used statistical 
physics~\cite{newman2006structure}.

Statistical physics tools have aided in formulating these microscopic models, 
which are simple enough yet rich in terms of socio-economic ingredients. Toy 
models help in understanding the basic mechanism at play, and demonstrate the 
crucial elements that are responsible for the emergent distributions
of income and wealth. A variety of models, ranging from zero-intelligence 
variants to the more complex agent based models (including those incorporating 
game theory) have been proposed over years and are found to be successful 
in interpreting the empirical results~\cite{chakrabarti2013econophysics}.
Simple modeling is also effective in understanding how entropy maximization 
produce distributions which are dominated by exponentials, and also explaining 
the reasons for aggregation at the high range of wealth, including the power law 
Pareto tail~\cite{yakovenko2009colloquium,chakrabarti2013econophysics}.

%%%%%%%%%%%%%%%%%%%%%%%%%%%%%%%%%%
\section{Summary and discussions}
\label{sec:summary}
%%%%%%%%%%%%%%%%%%%%%%%%%%%%%%%%%
Social inequalities are manifested in several forms, and are recorded well in 
history, being the reason of unrest, crisis, wars and revolutions.
Traditionally a subject of study of social sciences, though scholars from 
different fields have been investigating the  causes and effects from a 
sociological perspective, and trying to understand its consequence on the 
prevailing economic system. Reality is not as simple and  pointing out the 
causes and the effects are
much more complex.

Imagine a the world which is very equal, where it would have been difficult 
to compare the extremes, differentiate the good from the bad, hardly any 
leadership people will look up to, will lack stable ruling governments if there 
were almost equal number of political competitors, etc.

The recent concern about the increase of inequality in income and wealth,
as pointed out from different measurements~\cite{piketty2014capital} has 
renewed the interest on this topic among the leading social scientists across 
the globe. Society always had classes, and climbing up and down the social 
ladder~\cite{mervis2014tracking,bardoscia2013social} is quite difficult to 
track, until recent surveys which provide some insight into the dynamics.
Several deeper and important issues of our society~\cite{neckerman2004social} 
still need attention in terms of inequality research,
and this can only be achieved by uncovering hidden patterns on further 
analysis of the available data. 

Measurement of inequalities in society can be as simple as measuring measuring 
zeroth order quantities as Gini index, to finding exact probability
distributions. The complexity of the underlying problems have inspired 
researchers to propose multi-dimensional inequality 
indices~\cite{sen1992inequality}, which serve well in explaining a lot
of factors in a compact form.

As physicists, our interests are mostly concentrated on subjects which are 
amenable to modeling using macroscopic or microscopic frameworks. Tools 
of statistical physics can very well explain the emergence of broad 
distributions which are signatures of inequalities. The literature
already developed, contains serious attempts to understand socio-economic 
phenomena, under \textit{Econophysics}
and \textit{Sociophysics}~\cite{chakrabarti2007econophysics}.
The physics perspective brings  alternative ideas and a fresh outlook
compared to the traditional approach taken by social scientists,
and is reflected in the increasing collaborations between researchers across 
disciplines~\cite{lazer09}.

%%%%%%%%%%%%%%%%%%%%%%%%%%%%%%%%%%%%%%%%%%%
% \appendix
% %%%%%%%%%%%%%%%%%%%%%%%%%%%%%%%%%%%%%%%%%%%%%%
% \section{Derivation of overlap function:  A dynamical approach}
% \label{sec:AppA}
% %%%%%%%%%%%%%%%%%%%%%%%%%%%%%%%%%%%%%%%%% 
% In this appendix, we show the derivation 
% %%%%%%%%%%%%%%%%%%%
%%%%%%%%%%%%%%%%%%%%%%%%%%%%%%%%%%%%%%%%%%%%%%%%%%%%%%%%%%%%%%%%%%%%%%
\begin{acknowledgments}
%%%%%%%%%%%%%%%%%%%%%%%%%%%%%%%%%%%%%%%%%%%%%%%%%%%%%%%%%%%%%%%%%%%%%% 
A.C. thanks V.M. Yakovenko for discussions, in the context of a more extensive 
material which is being written up.
A.C. and B.K.C. acknowledges support from B.K.C.'s J.~C.~Bose Fellowship and  
Research Grant.
J.I. was financially supported by Grant-in-Aid for Scientific Research (C) of
Japan Society for the Promotion of Science (JSPS) No. 2533027803 and 
Grant-in-Aid for Scientific Research (B) of 26282089, 
Grant-in-Aid for Scientific Research on Innovative Area No. 2512001313. 
He also thanks Saha Institute of Nuclear Physics for their hospitality 
during his stay in Kolkata.
\end{acknowledgments}

%%%%%%%%%%%%%%%%%%%%%%%%%%%%%%%%%%%%%%%%%%%%
\bibliographystyle{unsrt}
\bibliography{kindex.bib}

\end{document}